\documentclass{aa}
\usepackage{graphicx,epstopdf}
\usepackage{hyperref}

\usepackage{txfonts}

\def\kmps{km\,s$^{-1}$}
\def\fergs{erg\,cm$^{-2}$\,s$^{-1}$}

\begin{document}

\title{Identification of 3XMM J000511.8+634018 as a new polar at $P_{\rm orb}=133.5$ min -- is it inside or outside the period gap?
}
\titlerunning{New polar 3XMM J000511.8+634018}

\author{A.D.~Schwope\inst{1}\and
H.~ Worpel\inst{1}\and
N.A.~Webb\inst{2}\and
F.~Koliopanos\inst{2}\and 
S.~Guillot\inst{2}
}
\institute{Leibniz-Institut f\"ur Astrophysik Potsdam (AIP), An der Sternwarte 16, 14482 Potsdam, Germany \\
\email{aschwope@aip.de}
\and
IRAP, Universit\'e de Toulouse, CNRS, UPS, CNES, 9 Avenue du
Colonel Roche, BP 44346, 31028 Toulouse Cedex 4, France
}

\authorrunning{Schwope et al.}
\date{}


\abstract
{}
{We aimed to identify the variable X-ray source 3XMM\,J000511.8+634018, which was serendipitously discovered through routine inspections while the 3XMM catalogue was compiled.}
{We analysed the archival XMM-Newton observation of the source, obtained BUSCA photometry in three colours, and performed optical spectroscopy with the LBT. These data were supplemented by archival observations from the Zwicky Transient Facility.}
{Based on its optical and X-ray properties, 3XMM J000511.8+634018 is classified as a magnetic cataclysmic variable, or polar. The flux is modulated with a period of 2.22\,hr ($8009.1\pm0.2$\,s), which we identify with the orbital period. The bright phases are highly variable in X-ray luminosity from one cycle to the next. The source shows a thermal plasma spectrum typical of polars without evidence of a luminous soft blackbody-like component.  It is non-eclipsing and displays one-pole accretion. The X-ray and BUSCA light curves show a stream absorption dip, 
which suggests an inclination $50\degr < i < 75\degr$. The phasing of this feature, which occurs at the end of the bright phase, requires a somewhat special accretion geometry with a stream running far around the white dwarf before it is magnetically channelled. The period of this polar falls within the period gap of the cataclysmic variables (2.15-3.18 h), but appears to fall just below the minimum period when only polars are considered.
}
{}
\maketitle

\section{Introduction}

Polars are cataclysmic variables (CVs) in which a strongly magnetic white dwarf, called the primary, accretes matter through Roche-lobe overflow  from a late-type main-sequence companion star, called the donor. Because of the strong magnetic field \citep[$\sim 10-200$\,MG;][]{ferrario+15} no accretion disc forms; instead, in-falling gas initially travels along ballistic trajectories and later along the magnetic field lines and is deposited onto a comparatively small region near the magnetic pole, or poles, of the primary.

When the gas approaches the surface of the white dwarf, a standing shock forms and thermal plasma radiation is emitted isotropically at X-ray energies. In addition, the plasma cools through cyclotron radiation at optical and neighbouring wavelengths.  The plasma emission observed at many wavelengths is among the most important observational signatures of a polar. It is typically detected with a temperature of a few tens of keV, often modulated at the orbital period of the system (a few hours). Their brightness, distinctive X-ray and optical light curves, and hard X-ray spectra make serendipitous discoveries relatively common \citep[e.g.][]{vogel+08a,ramsay+09,webb+18}.

Many polars have been discovered through their pronounced soft X-ray emission in the ROSAT all-sky survey. The excess of soft blackbody-like X-ray emission of reprocessed origin over the hard primary emission from shocked plasma led to the infamous soft X-ray puzzle in AM Herculis stars \citep{beuermann_schwope94}. A more recent  account of it was given by \cite{RamsayCropper2004} based on XMM-Newton follow-up. The authors found a soft excess in only about 20\% of their sources. As a novelty, a considerable fraction of objects among their sources did not show a soft-component at all.

The evolutionary history of CVs is written in their orbital period distribution \citep[e.g.][and references therein]{knigge+11}. Thus the orbital period is the first fundamental piece of information to be determined from photometric or spectroscopic observations. The period distribution of non-magnetic CVs shows the famous period gap between 2 and 3 hours \citep[exact values determined by ][as 2.15 and 3.18 hours, respectively]{knigge+11}. The gap seems to be existent in magnetic CVs as well, but was found to be less prominent \citep{pretorius+13}, lending evidence to a yet to be formulated modified magnetic braking prescription. An ab initio theory for braking and population synthesis for magnetic CVs is lacking \citep{pala+19}. Today, we know about 140 polars \citep[][and occasional additions since they stopped updating their catalogue]{ritter+kolb03}. Most polars are observed at short orbital periods (110 objects below $P_{\rm orb} = 170$ min); polars with periods above the gap are rare. Adding further members is thus valuable to further substantiate the existence of a gap in the period distribution of polars and to constrain its properties.

Screening the standard XMM-Newton data products of the 3XMM catalogue \citep{RosenEtAl2016} yielded the object 3XMM J000511.8+634018, hereafter J0005, observed near the bright star HD~108 \citep{NazeEtAl2004}, with a light curve that was 100\% variable and strongly reminiscent of a compact binary. We reanalysed these data to conclusively identify the object. Our archival data are complemented by 2.5 hours of multi-colour optical observations at Calar Alto performed in 2018 October, optical spectroscopy performed at the Large Binocular Telescope (LBT) in 2019 November, and several months of monitoring by the Zwicky Transient Facility \citep[ZTF;][]{MasciEtAl2019}.

The source is listed in Data Release 2 of the Gaia catalogue \citep{Gaia2016, Gaia2018}; its mean magnitude was $m_g=20.12$ and its  parallax negative, $ \pi = -0.10 \pm  0.75$. Hence, the distance is only poorly constrained to be $2.8^{+2.2}_{-1.3}$\,kpc \citep{Bailer-JonesEtAl2018}.

\section{X-ray observations}

We retrieved the XMM-Newton observation of J0005 (obsid 0109120101, 2002 August 21). The duration was 36.7\,ks. The field was observed with the EPIC-MOS instruments in full-frame mode and EPIC-pn in extended full-frame mode. All exposures used the thick filter. The actual target of the observation, HD~108, is bright enough to cause damage to the Optical Monitor on XMM-Newton, so no optical or ultraviolet data were taken. We found no other archival observations of J0005 by other X-ray observatories. J0005 had a mean EPIC count-rate (mean of the three X-ray cameras) of $0.0335\pm0.0015$\,s$^{-1}$ , yielding $764\pm33$ source photons between 0.2 and 12 keV.

\begin{figure}
    \includegraphics[width=0.49\textwidth]{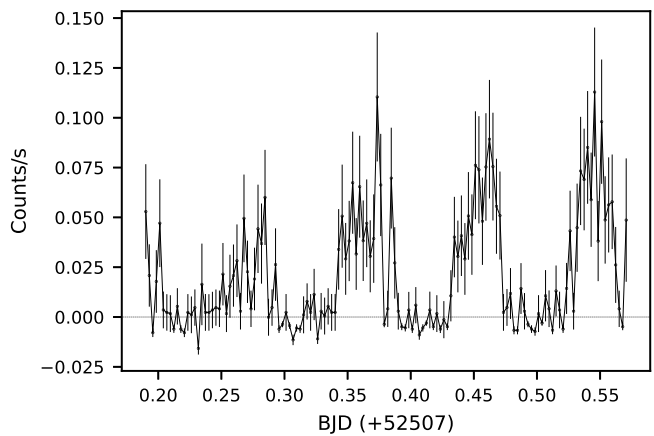}
    \caption{EPIC-$pn$ X-ray light curve (0.2-10 keV) of J0005, background corrected and binned to time bins of 240\,s.}
    \label{f:xlc}
\end{figure}

\begin{figure}
    \includegraphics[width=0.48\textwidth]{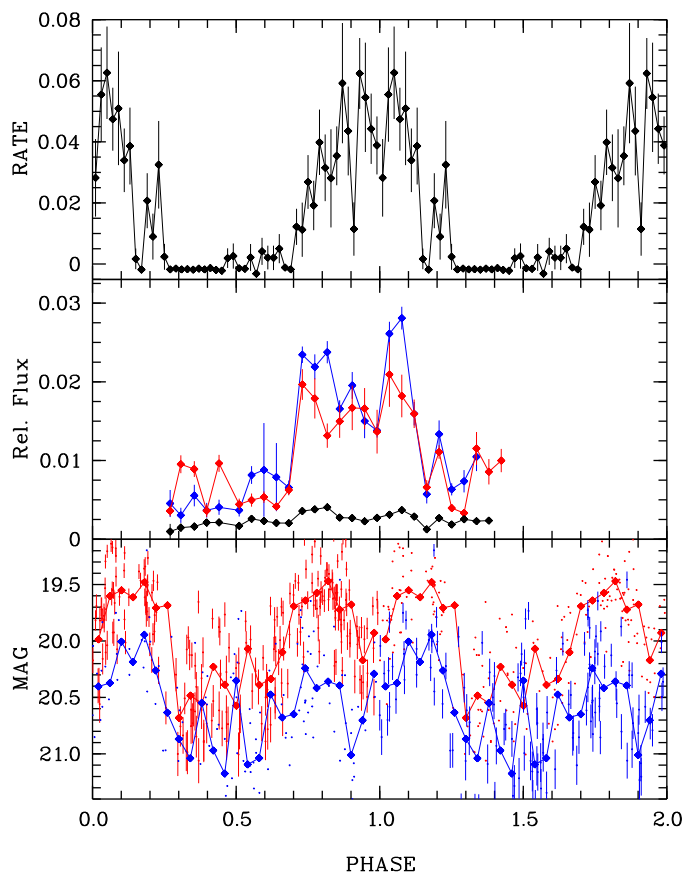}
    \caption{Phase-folded X-ray (top panel, 50 phase bins), BUSCA optical (middle panel), and ZTF optical (bottom panel, 25 phase bins) light curves of 3XMM J000511.8+634018. Phase zero was placed in the centre of the X-ray bright phase, BUSCA data shown in blue, red, and black for the blue, red, and UV filters were adjusted in phase so that the absorption dips coincide. We show the flux ratio with regard to the photometric comparison star. ZTF data through $g$- and $r$-filters shown in blue and red were shifted to approximately coincide with the X-ray data. In the bottom panel both the original (small dots) and the phase-binned (larger dots connected by lines) data are shown. In the top and bottom panels two cycles are plotted for clarity, but error bars for ZTF data are shown only for one cycle for better visibility.
    \label{f:xmm_phasefold}
}
\end{figure}

The X-ray light curve of the source (see Figure \ref{f:xlc}) shows obvious bright and faint phases. This type of light curve is typical for a polar and indicates that an accretion region rotates in and out of view. To derive the period of the modulation, we performed an H-test \citep{deJagerEtAl1989} on the photon arrival times and obtained an orbital period of $7,988\pm 57$\,s, with the $1\sigma$ uncertainty determined with the bootstrap method. This period is consistent with the long-term optical period found in ZTF data (see Section~\ref{s:opt}). The EPIC-$pn$ light curve folded on this period, covering $\sim 4.2$ cycles, is shown in Figure \ref{f:xmm_phasefold} (top panel). The bright phase lasts 0.55 phase units, and there is a precursor to the genuine bright phase lasting about 0.2 phase units. This precursor and the bright phase seem to be separated by a short phase interval ($\leq 0.05$ phase units) with zero flux. During the remaining part of the faint phase, the flux is consistent with zero. The shape of the light curve suggests that we see accretion at one dominant pole. Whether the short precursor is to be associated with a second accretion region or is related to the main region cannot be answered based on the few photons that were counted in the phase interval. The low photon count also prevents us from determining whether the precursor is present in all cycles, but to the eye, it appears more pronounced in the first two faint phases of the X-ray light curve.

\begin{figure}[t]
    \includegraphics[width=0.5\textwidth]{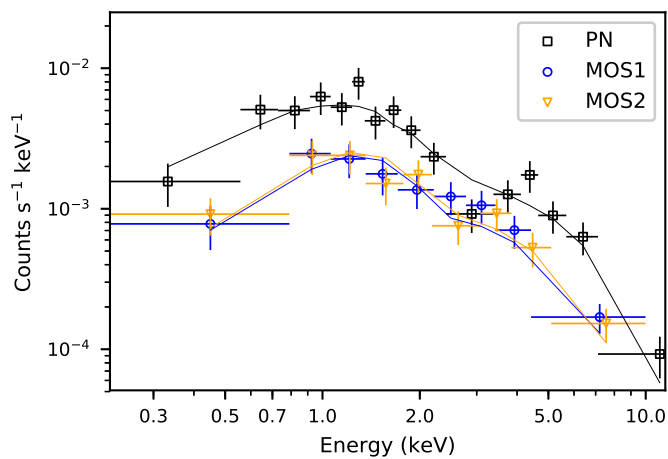}
    \caption{X-ray spectral fit of the average bright phase of J0005. The data are shown as points, and the model fit {\tt TBabs*apec} are represented as solid lines.}
    \label{f:xmmspecfit}
\end{figure}

The luminosity of the bright phases is evidently variable, with a continuous increase in brightness over the four cycles observed by XMM-Newton. A distinct dip also occurs towards the end of the bright phase. This feature is present in all bright phases, on five occasions\footnote{The feature is apparently missing at BJD 52507.46 in Fig.~\ref{f:xlc}, which is only due to the somewhat coarse binning that was used to produce the figure to avoid confusion.}, as well as in the photometric data obtained with BUSCA (see Section \ref{s:opt}). The length of the feature was measured in a phase-folded X-ray light curve and determined to last approximately 300\,s (0.038 phase units). During that time interval no X-ray photons were measured from the source. 

The bright phases of J0005 show a steadily increasing luminosity. A similar phenomenon was recently reported in \cite{ReaEtAl2017} for the intermediate or asynchronous polar RX~J0838$-$2827, which the authors suggested was the beat period between the orbital period and the white dwarf spin period. We tested this possibility for J0005 as follows. It is evident from Figure \ref{f:xlc} that if a beat period is present, it must be more than twice as long as the XMM-Newton observation. Thus, any additional period must be longer than about 7,200\,s or shorter than 8,900\,s in order for the beat period to exceed $2\times 36.7$\,ks. However, the ZTF data showed no additional periods in this range except for the one-day aliases of the 8009.1\,s signal.

There are not enough photons in the observation to extract spectra for each individual bright phase, so we proceeded by assuming a constant spectral shape with variable normalisation and extracted a mean bright phase spectrum. We grouped it to a minimum of 16 counts per bin, and fit it between 0.2 and 10.0\,keV to an absorbed plasma model \citep[{\tt TBabs*apec} in XSPEC;][]{Arnaud1996}. All three spectra (EPIC-$pn$, MOS1, and MOS2) were fit jointly. We obtained an acceptable fit ($\chi^2_\nu=0.91$ for 31 degrees of freedom) with this model. The temperature was not well constrained, but a temperature lower than 7.5\,keV is ruled out at the 95\% confidence level. The temperature was then fixed at 10 keV, a typical value found among the polars that is difficult to measure precisely, however, given the limited spectral window by XMM-Newton. Furthermore, the observed spectra are subject to intrinsic absorption and reflection, which results in spectral hardening \citep{mukai2017}, so that an unconstrained fit tends to run to the maximum temperature allowed by the model.

The best-fit hydrogen column density of the absorber depends slightly on the (highly uncertain) plasma temperature. For a high temperature of 64\,keV, we obtain $1.09^{+0.03}_{-0.02}\times 10^{21}$\,cm$^{-2}$, with the $1\sigma$ uncertainty determined with the {\tt steppar} command of Xspec. If the temperature is instead fixed at 10\,keV, we obtain $n_H=1.49^{+0.30}_{-0.27}\times 10^{21}$\,cm$^{-2}$. How do these values compare with the literature? The \cite{BekhtiEtAl2016} gives $6.1\times 10^{21}$\,cm$^{-2}$ for this sky region, which is not consistent with our result. A more recent 3D study by \cite{GreenEtAl2019} gives $1.6^{+0.95}_{-0.15}\times 10^{21}$\,cm$^{-2}$ for the coordinates and the Gaia distance range of $2.8^{+2.2}_{-1.3}$\,kpc of J0005. This value agrees well enough with ours, and seems to suggest that J0005 is at the nearer end of its plausible Gaia distance range, that the plasma temperature is $\sim 10$\,keV, and that the system contains little intrinsic absorption, but the data quality is not good enough to make any definitive statement.

A soft blackbody-like component was commonly found in the spectra of polars and some intermediate polars in the ROSAT data \citep[e.g.][] {KuijpersPringle1982, RamsayCropper2004}. We note that there was no need to include such a component here. The X-ray fit is shown in Figure \ref{f:xmmspecfit}. We also found no strong evidence of a fluorescent iron line at 6.4\,keV: formally, the fit is improved slightly by including it, but its normalisation is consistent with zero at the $1.25\sigma$ level, and the $1\sigma$ upper limit of its equivalent width is $0.9$\,keV.

The absorbed flux in the 0.2--10.0\,keV energy range was $(1.55\pm0.07)\times 10^{-13}$\,erg\,s$^{-1}$\,cm$^{-2}$ ($1.71\pm0.08$ unabsorbed). The luminosity in the same energy range is therefore $1.5^{+1.6}_{-1.0}\times 10^{32}$\,erg\,s$^{-1}$ ($1.7^{+1.8}_{-1.1}\times 10^{32}$\,erg\,s$^{-1}$ unabsorbed). The large uncertainty is due to the poorly known distance to the system, but the luminosity value is consistent with it being a polar. \cite{pretorius+13} determined the X-ray luminosity for all polars in the ROSAT Bright survey \citep[RBS, ][]{schwope+02} and found a mean of $1.6 \times 10^{32}$\,erg\,s$^{-1}$.

\begin{figure}
    \includegraphics[width=0.5\textwidth]{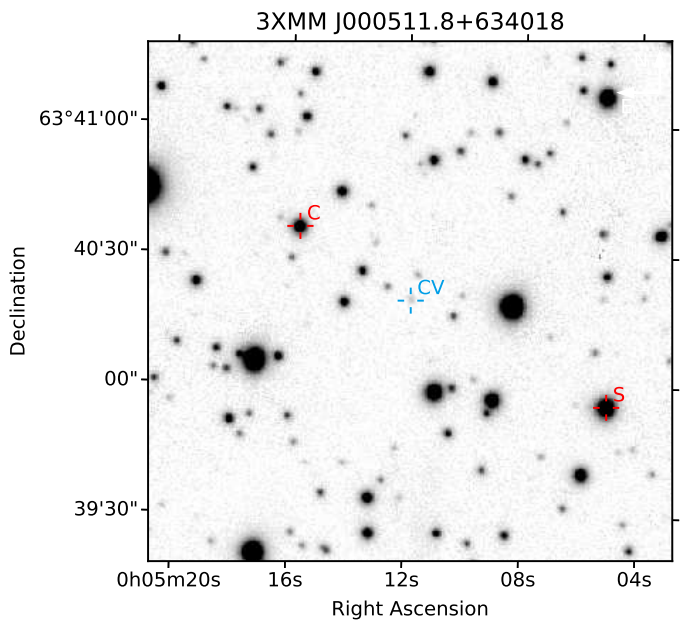}
    \caption{Optical finding chart (PanSTARRS $g$) with the target marked by a cross, and the photometric and spectroscopic reference stars with letters C and S, respectively. The chart has a size of $2\times2$\,arcmin. North is at the top and east to the left.
    \label{f:fc}}
\end{figure}

\section{Optical photometric observations}
\label{s:opt}

We observed J0005 for 2.5 hours on 2018 Oct 17 with the Bonn University Simultaneous CAmera \citep[BUSCA, ][]{reif+00} instrument on the 2.2\,m telescope at Calar Alto. We obtained 27 images, of which the first had 90\,s exposure and the rest were 300\,s. BUSCA is capable of observing in four colours simultaneously, but only three channels were working during our observation. We used the red, blue, and ultraviolet channels with broad UV$\_$B, B, and R filters\footnote{https://www.caha.es/guijarro/BUSCA/caracter.html}, respectively, without the near-infrared. To avoid excessive readout time, we only used the central $1024\times 1024$ pixels of the CCDs, which reduced the readout time to 45 seconds.

We used AstroImageJ \citep{CollinsEtAl2017} to perform bias and flat-field corrections, to align the images, and to extract differential photometry against a brighter star (00:05:15.7, +63:40:36) in the field. In the finding chart in Fig.~\ref{f:fc} the target and the reference star are marked with a cross and the letter 'C'. To correct the time stamps to the Solar System barycenter, we used Astropy \citep{Astropy2018}. Unfortunately the field is very crowded: the source lies in the Galactic plane, and there is a strip of bad pixel values, caused by the overexposure of a bright nearby star that partially covers the target. To obtain the photometry of J0005, we fit the pixel values of our target with a two-dimensional Gaussian within a rectangular region extending to the strip of bad pixels and obtained the brightness by computing its integral. We repeated this for the reference star.

The light curves of this observation are shown in the middle panel of Figure \ref{f:xmm_phasefold}. Time was converted into phase using a period of 8009.1\,s and an arbitrary zero-point. The BUSCA data show a bright phase with a dip towards the end in all three bands, very reminiscent of the X-ray data. 
The similarity of the overall shape of the X-ray and BUSCA light curves (width, dip, etc.) suggest a similar origin or region for the optical and X-ray emission near one main accretion pole.
For the presentation in Fig.~\ref{f:xmm_phasefold}, the BUSCA data were shifted so that the dips at the end of the bright phase coincide.
The comparison star is in the PanSTARRS catalogue \citep{ChambersEtAl2016} at $m_g=16.66$, $m_r=15.5$. When we consider these to be roughly equivalent to BUSCA blue and red filters, respectively, we estimate the magnitude of J0005 to be about 20 and 21 in red and blue respectively during bright phase, and 22 and 21 in the faint phase. There are no archival UV observations, therefore we cannot estimate the magnitude in the UV.

The ZTF \citep{MasciEtAl2019} lists 355 observations of J0005 in January 2020. These were taken from 2018 May 6 to 2019 June 29 with exposure times of 30\,s with time differences between 0.03 and 50 days within one observing season and a median time difference of 1.9 and 1.1 days among the $g$- and $r$-band data, respectively. Of the 335 photometric measurements 128 were obtained with the $g$-filter and 227 with the $r$-filter. While ZTF observed the field of J0005 rather frequently at times, our BUSCA observation falls almost in the middle of a data gap lasting $\sim$20 days in the ZTF so that no direct comparison of the brightness levels is possible (given the short timescales on which the transition between high and low states may occur). The ZTF shows the source magnitude ranging between 21.6-19.2 in $g$ and 21.1-18.7 in $r$. When the data are inspected in the original time sequence, a clear trend is recognised. In July 2018, the minimum $g$-magnitude was 21.6, and in June 2019, the minimum was about 1 mag brighter, indicating a switch between a low (or reduced) and a high accretion state. The source was always variable by $1-1.5$ magnitudes. The $r$ band shows a similar trend in the overall brightness between a minimum around $r=21$ in mid-2018 and a minimum around 20 in mid-2019. The large scatter of data is naturally explained by the pronounced short-term variability on the timescale of 2 hours as seen in the BUSCA and the XMM-Newton data. We thus downloaded the data, which already have heliocentric timestamps, and sought periodicities using the analysis-of-variance method \citep[AoV; e.g. ][]{Schwarzenberg-Czerny1989} on the datasets from the two filters individually. We did so by searching periodicities directly in the observed data thus ignoring the overall brightness trends. 

There was an obvious signal at around 8009\,s. For the $r$ and $g$ filters we obtained periods of $8,009.1\pm0.2$\,s and $8,009.5\pm 66.2$\,s, with $1\sigma$ uncertainties calculated using the bootstrap method \citep{Efron1979}. For the rest of the analysis, we take the period to be 8,009.1\,s.

The ZTF light curves in $r$ and $g$ filters, folded on the orbital period using 25 phase bins, are shown in the bottom panel of Figure \ref{f:xmm_phasefold}. The data are plotted as small dots and the phase-averaged data as solid lines. There is no dip at the end of the bright phase, as there was in the BUSCA data, but both sets of optical observations show a period of decreased luminosity in the middle of the bright phase.

\begin{figure}
    \includegraphics[width=0.33\textwidth,angle=-90]{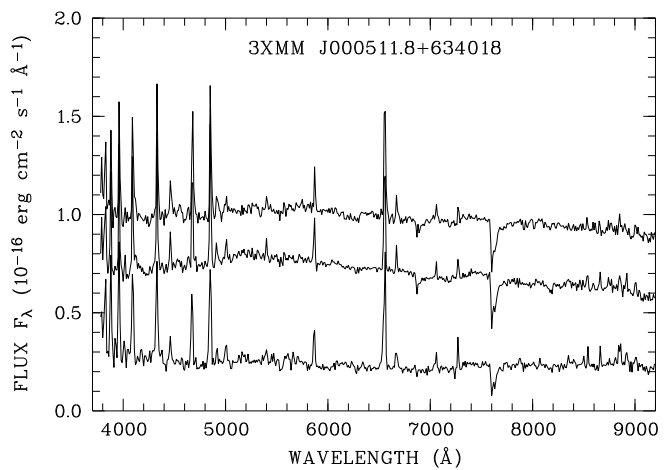}
\caption{LBT spectra obtained on 2019 November~24. The spectra are arranged in time sequence from bottom to top. Spectra 2 and 3 were shifted vertically by 0.3 and 0.6 flux units.
    \label{f:lbtover}}
\end{figure}

\begin{figure}
    \includegraphics[width=0.33\textwidth,angle=-90]{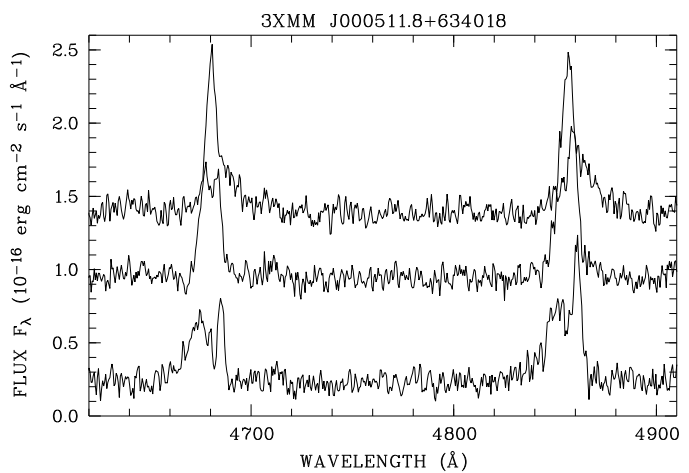}
  \caption{Detail of LBT spectra obtained on 2019 November~24, showing the variability of the H$\beta$ and the HeII 4686 emission lines. The spectra are arranged in time sequence from bottom to top. Spectra 2 and 3 were shifted vertically by 0.5 and 1.0 flux units.  \label{f:lbtdet}}
\end{figure}

\section{LBT spectroscopy}
J0005 was observed spectroscopically with the LBT, which was equipped with the two multi-object double spectrographs (MODS) on the night of 2019 November 24. For a description of MODS, see \cite{pogge+10}. Weather conditions were unstable during that night, but for the three spectra obtained with 600 s exposure each, the conditions were rather stable. The 1.2 arcsec segmented long slit allowed us to place a second object on the slit with the PanSTARRS designation PSO J000513.565+634025.950. This was used to correct for slit losses. A dispersion solution was found using arc lamp spectra that were obtained two days before the observations of the target star. The standard star Feige 110 was observed through a wide slit (5 arcsec) during the same night as our target.
The data were pre-reduced (bias- and flat-field correction) with the {\tt modsCCDRed} software\footnote{\url{http://www.astronomy.ohio-state.edu/MODS/Software/modsCCDRed/}} , while the wavelength calibration, the spectral extraction, the flux calibration and the photometric correction using the second object on the slit were performed with ESO-MIDAS.

Useful spectral information was extracted between 3780\,\AA\ and 9200\,\AA\ with a full width at half-maximum (FWHM) spectral resolution of 2.3\,\AA\ in the blue arms and 3.5\,\AA\ in the red arms of the two spectrographs (below and above the wavelength split at about 5500\,\AA\ given by the dichroic used). The data from MODS1 and MODS2 were obtained with a negligible timing offset so that the data from both spectrographs were averaged. We estimate that the finally achieved overall photometric accuracy of our spectra is about 20\%. 

The three spectra obtained with mid-exposure times at MJD 58811.251448, 58811.259682, and 58811.267996 are shown in Figs.~\ref{f:lbtover} and \ref{f:lbtdet} to illustrate the brightness distribution over the full spectral range covered in the former and to focus on the strong HeII\,4686 and H$\beta$ emission lines in the latter figure. The overall brightness in $g$ and $r$ filters for the three spectra was $g=20.5, 20.0, 20.1$, and $r= 20.1,19.5, 19.4$, respectively, therefore the source appeared to be brighter in November 2019 than in previous observations by about half a magnitude. Pronounced continuum variability was clearly present. The first spectrum was probably taken during the photometric faint phase, while the two following spectra were taken during the bright phase. While the faint-phase spectrum is flat in $F_\lambda$, the two bright-phase spectra display a distinct spectral maximum at 5500\,\AA.

\begin{table}
\caption{Emission line parameters determined in the mean LBT spectra\label{t:lines}}
\begin{tabular}{lrr}
\hline
Line & FWHM ($\AA $) & EW ($\AA $)\\
\hline
H$\epsilon$ & 11.6 & $-26.8$ \\
H$\delta$ & 10.9 & $-21.0$\\
H$\gamma$ & 11.6 & $-24.2$\\
HeI\,4471   & 14.0 & $-8.0$\\
HeII\,4686  & 12.0 & $-20.4$\\
H$\beta$  & 13.3 & $-28.0$\\
HeI\,4921   & 15.3 & $-4.1$\\
HeI\,5015   & 12.0 & $-3.8$\\
HeII\,5411  & 14.5 & $-3.2$\\
HeI\,5875   & 13.4 & $-10.7$\\
H$\alpha$   & 14.8 & $-34.9$\\ 
HeI\,6678   & 12.4 & $-5.2$\\
HeI\,7065   &  9.7 & $-3.5$\\
CaII\,8498$\dagger$  & 10.9 & $-1.9$\\
CaII\,8542$\dagger$  & 15.6 & $-3.7$\\
CaII\,8662$\dagger$  &  8.9 & $-3.0$\\
\hline
\end{tabular}
\newline $\dagger$ the CaII triplet lines might be blended with P10, P12, and P13, respectively
\end{table}

The emission lines have strongly variable profiles. A narrow and a broad line can be discerned in the first two spectra. The full width at zero intensity (FWZI) of HeII\,4686 is around 3000\,\kmps\ in the mean of the three spectra. A few further emission line parameters of prominent H, He, and Ca lines lines are listed in Table~\ref{t:lines}. The values given there are based on a Gaussian fit to the emission lines in the mean of the three spectra. They are indicative only because such values are seen to be strongly phase-dependent in other systems \citep[e.g.~][]{schwope+97b}. The FWHM of the narrow emission lines of HeII\,4686 is 3.6\,\AA, that of the Balmer lines is larger at $5-6$\,\AA, respectively. The Bowen blend CIII/NIII is weakly detected. Its presence indicates a photo-ionising source of radiation. The behaviour of the continuum, its shape and variability, and the composition of the emission line spectrum, the line parameters, the high ratio of the equivalent widths $EW(HeII\,4868)/EW(H\beta)\sim0.7$: all these properties are very reminiscent of an origin in a polar \citep[e.g.][]{schwope+97b,schwope+00,szkody+11}. The narrow emission line is naturally identified with emission from the irradiated side of the donor star, while the broad emission line originates in streaming matter between the two stars. 

\section{Discussion and conclusion}

The source 3XMM~J000511.8+634018 is almost certainly a polar cataclysmic variable. Only one piece of evidence may be missing: a unique detection of magnetism through Zeeman absorption or cyclotron emission lines, or more generally, through the detection of a strongly polarimetric signal. Otherwise, J0005 shows all the hallmarks of a polar from the X-ray and optical data presented here. Evidence in the X-ray regime is derived from the timing and spectral behaviour, and from its luminosity. 
The thermal X-ray spectrum is typical for this class, as are the characteristic bright and faint phases that arise when the accreting pole rotates in and out of view. The X-ray and optical brightness are modulated with a period of 2.2 hours, which in the first instance indicates the spin period of the white dwarf and for most ($\sim$95\%) objects of this class also the orbital period of the binary. At this orbital period, the object is placed at the bottom end of the orbital period gap of non-magnetic CVs. The period distribution and the location of the new source therein are further discussed below.

Evidence in the optical regime for a polar classification is derived from the shape of the light curves and the identification spectra. Optical light curves are modulated on the same periodicity as the X-ray light curve. The double-humped shape of the optical light curve is reminiscent of cyclotron beaming \citep[see e.g.][for a similar light curve]{schwope+03}. The cyclotron spectrum peaks at around 5500\,\AA. This wavelength approximately amrks the transition from the optically thick part of the cyclotron spectrum at long wavelengths to the optically thin part at short wavelengths. Because our spectra are slightly noisy and most of the observed part of the spectra is in the optically thick regime, we did not find cyclotron harmonic humps that would have allowed a direct measurement of the magnetic field strength. However, other polars with similar spectral appearance had measured magnetic field strength between $B=20-25$\,MG 
\citep[e.g.~V834 Cen and MR Ser,][respectively]{schwope+beuermann90, schwope+93b}.

The emission line pattern observed with the LBT can be explained through its similarity to objects that were observed with full phase coverage and with high time and spectral resolution \citep[see e.g.][for the case of HU Aqr, their Figure 4]{schwope+97b}. The merging broad and narrow emission lines suggest that the LBT spectra were obtained between binary phases 0.5 and 0.7. At phase $\sim$0.5 (first spectrum), we typically see the highest blue-shift of the broad component, which originates from streaming matter, and the narrow emission line (NEL) has zero velocity. In the following spectra, the NEL becomes blue-shifted while the observer sees the streaming matter more from the side at lower velocities. At phase $\sim$0.7, the line components can no longer be uniquely distinguished. When the streaming matter is seen from the side, the projected velocities in the stream come close to zero, thus mimicking a true NEL. In sum, the emission line variability leaves little doubt about the polar nature of J0005.

The luminosity dip at the end of the bright phase is evident in X-rays and all three BUSCA light curves, but it is not visible in the ZTF data. X-ray and optical absorption signatures may arise from an eclipse of the white dwarf by the donor star and by intervening matter in the accretion flow. The lack of dips in the ZTF data favours the latter explanation, although the absence of dips in ZTF data should not be over-interpreted. The integration times in the ZTF is only 30 s; dips are known to be highly variable in phase, they are not monolithic, but may show internal structure. Hence the short ZTF exposures may simply have missed the dip at those times. Such stream dips are seen often in polars, but J0005 is unusual because the dip occurs at the end of the bright phase rather than near the middle. This means that the accreting matter runs far around the white dwarf in the orbital plane before it is threaded onto magnetic field lines and then moves back towards the binary meridian. Because there is no evidence for this dip in the ZTF data,  this might mean that the described accretion geometry might at times become even more extreme so that the stream never crosses the line of sight towards the accretion spot, but appears later in phase. This could be understood as an indication of asynchronous rotation, but remains speculative for now. 

The X-ray spectral fit shows a somewhat low hydrogen column density for a source near the Galactic plane (galactic coordinates of J0005 are $l^{II}=117.83\degr$ and $b^{II}=1.26$\degr). We therefore tentatively infer that J0005 is at the nearer end of the very uncertain distance measurement ($2.8^{+2.2}_{-1.3}$\,kpc). A smaller distance would give an X-ray luminosity of about $5\times 10^{31}$\,erg\,s$^{-1}$, similar to that of other period-gap polars \citep[e.g.~2PBC~J0658.0-1746; ][]{BernardiniEtAl2019}.

The X-ray spectrum is not of very high quality, but at the signal-to-noise ratio we achieved, we did not find evidence for the ubiquitous luminous soft-radiation component found in all polars prior to the XMM-Newton era. The current finding (or non-finding of a soft component) adds to several other non-detections in XMM-Newton discovered polars \citep[see][for two recent examples and further references therein]{webb+18}. The intrinsic X-ray spectrum of a parent polar sample remains unknown.

The companion does not eclipse the accreting white dwarf. The system seems to have one dominant pole. This primary pole is only visible a little over half the spin phase, and a stream dip was observed at X-ray and optical wavelengths, which suggests an inclination angle in the range $50\degr < i < 75\degr$.

The bright phases of J0005 show significant variability from one cycle to the next in X-rays, but there is no evidence for a beat period such as might be seen in a slightly asynchronous polar. Only five asynchronous polars are known. J0005 might be another example, but repeated spectroscopic observations with full phase coverage are required to establish a precise value of the orbital period. Further photometric monitoring is likewise required to assess whether the white dwarf rotates truly synchronously.

\begin{figure}[t]
    \includegraphics[width=0.33\textwidth,angle=-90]{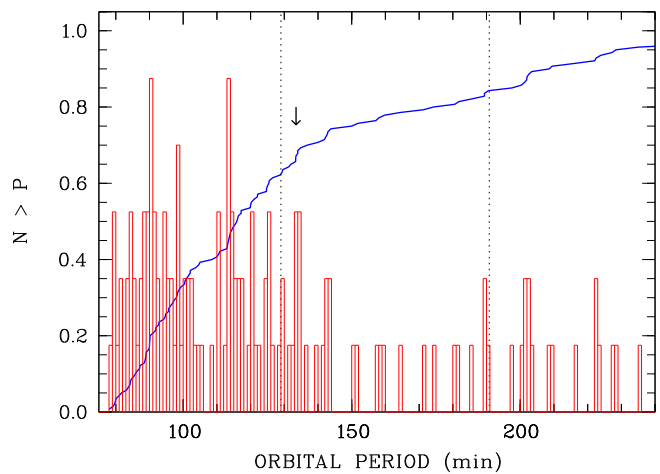}
\caption{Normalised cumulative (blue) and binned (red) orbital period distribution of the polars. The location of J0005 is indicated with an arrow. the bin width is 1\,min, the highest bin has five members. The ordinate scale is valid for the cumulative distribution. The dashed lines indicate the lower and upper bounds of the period gap as determined by \cite{knigge+11}
\label{f:pdist}}
\end{figure}

When we use $f_x=1.7\times 10^{-13}$\,\fergs\ and $g=20$ (maximum brightness) to compute the ratio of X-ray to optical flux, the achieved value of $\sim$0.6 compares well with other X-ray selected CVs \citep[see e.g.~][their figure 11]{comparat+19}. This does not further constrain the sub-type, however.

The final part of this paper is devoted to a discussion of the orbital period distribution of the polars, motivated by our observation that the period of J0005 seems to lie within the famous period gap between 2 and 3 hours. The period distribution encodes the birth rate and the secular evolution of the CVs and is thus of fundamental relevance for an understanding of the evolution of the close binaries. The existence of the period gap and its width are usually explained by the disrupted magnetic braking model \citep[DBM, ][]{rappaport+83}. Above the gap, the evolution of the binaries is driven mainly by magnetic braking. The upper bound $P_{+}$ marks the period (hence the mass of the secondary) where the donor star becomes fully convective and the large-scale ordered magnetic field is assumed to break down.  The evolution through the gap without mass transfer and at short periods is driven mainly by gravitational radiation. 

The evolution of the CVs through the properties of their donor stars 
was intensively studied by \cite{knigge+11}, who also 
determined the locations of the lower and upper bounds of the gap as $P_{-} = 2.15$ hours and $P_{+}=3.18$\,hours. Most studies, including that of \cite{knigge+11}, describe the evolution of non-magnetic CVs. The magnetic CVs seem to behave differently. \cite{pretorius+13} noted the presence of a period gap that they found to be less pronounced than that of the non-magnetic CVs. \cite{pala+19} noted an unexpected high fraction of 36\% of magnetic CVs in their volume-limited sample of 42 objects and concluded that future population synthesis models need to take magnetic objects into account. A first attempt in this direction was recently presented by \citet{belloni+20}.

Starting with the last edition of the catalogue of \cite{ritter+kolb03} and adding occasional reports of new polars in the literature, we count 140 objects with measured orbital periods. The periods range from 79 minutes to 480 minutes. Their distribution is shown in Fig.~\ref{f:pdist} in a binned and an un-binned (cumulative) form. The location of J0005 is indicated by an arrow, and the bounds of the gap as determined by \cite{knigge+11} are shown as well. The distribution shows a gap, but its lower bound $P_{-}$ is not found at 2.15 hours (129 min), but at a longer period. The cumulative distribution shows two jumps after which a flattening of the distribution is seen, one at 137 min, the other at 145 min. Hence we have two candidates for $P_{-}$ in the period distribution of the polars. At a period of 133.5 min, J0005 lies very close to the shorter of the two periods. Given the small number of polars at long orbital periods, it is difficult if not impossible to determine a precise value of $P_{+}$ for the polars. The data are compatible with the same value as for the non-magnetic CVs. 

The DBM assumes that the donor stars are driven out of equilibrium above the gap (too large radii for the given mass compared to main-sequence stars of the same mass) and relax to their main-sequence radius below the Roche surface when the dominant angular momentum loss mechanism brakes down. This happens to the donors of all CVs, whether they are magnetic or non-magnetic. There is no reason to assume that $P_{+}$ is very different for magnetic and non-magnetic CVs. The degree of bloating of the donors, however, is expected to be different. The strong magnetic field of the white dwarf reduces the number of open field lines so that magnetic braking above the gap is expected to be less effective for magnetic than for non-magnetic objects \citep[cf.][]{li+95}. If braking is reduced, the mass loss above the gap is lower, the donors will be less eroded than their cousins in a non-magnetic CV. Consequently, the donors may regain contact to the Roche surface at longer orbital periods. Because the fraction of open field lines depends on the orientation of the magnetic axis in the white dwarf, the braking depends on the field geometry \citep{li_wickramasinghe98}. It may thus happen that no well-defined value of $P_{-}$ exists for polars because the strength of magnetic braking is different for each object. This scenario needs more objects for an observational confirmation that are ideally drawn from complete flux-limited or better volume-limited samples.

The prospects to achieve this over the next ten years are excellent.
The mean X-ray flux for the whole observation of the source in the soft energy band between 0.5 and 2.0 keV as listed in 3XMM was $f_{\rm x} = (1.89\pm0.16)\times 10^{-14}$\,ergs\,cm$^{-2}$\,s$^{-1}$. This is a factor 2 above the limit of the  eROSITA surveys \citep{merloni+12} that have just commenced taking data. Many of these sources may reasonably be expected to be detected within a volume of several kiloparsec in the years to come. The open question of the size of the period gap in polars may thus eventually be settled.

\begin{acknowledgements}
We thank an anonymous referee for helpful comments and the LBT observers A.~Becker, A.~Wittje, and J.T.~Schindler for their support at the telescope. We thank Jan Kurpas for assistance. 

This work was supported by the German DLR under contract 50 OR 1814. This research has made use of data obtained from the 3XMM XMM-Newton serendipitous source catalogue compiled by the 10 institutes of the XMM-Newton Survey Science Centre selected by ESA. 

This paper used data obtained with the MODS spectrographs built with funding from NSF grant AST-9987045 and the NSF Telescope System Instrumentation Program (TSIP), with additional funds from the Ohio Board of Regents and the Ohio State University Office of Research. 

The Pan-STARRS1 Surveys (PS1) have been made possible through contributions of the Institute for Astronomy, the University of Hawaii, the Pan-STARRS Project Office, the Max-Planck Society and its participating institutes, the Max Planck Institute for Astronomy, Heidelberg and the Max Planck Institute for Extraterrestrial Physics, Garching, The Johns Hopkins University, Durham University, the University of Edinburgh, Queen's University Belfast, the Harvard-Smithsonian Center for Astrophysics, the Las Cumbres Observatory Global Telescope Network Incorporated, the National Central University of Taiwan, the Space Telescope Science Institute, the National Aeronautics and Space Administration under Grant No. NNX08AR22G issued through the Planetary Science Division of the NASA Science Mission Directorate, the National Science Foundation under Grant No. AST-1238877, the University of Maryland, and Eotvos Lorand University (ELTE).

Based on observations obtained with the Samuel Oschin 48-inch Telescope at the Palomar Observatory as part of the Zwicky Transient Facility project. ZTF is supported by the National Science Foundation under Grant No. AST-1440341 and a collaboration including Caltech, IPAC, the Weizmann Institute for Science, the Oskar Klein Center at Stockholm University, the University of Maryland, the University of Washington, Deutsches Elektronen-Synchrotron and Humboldt University, Los Alamos National Laboratories, the TANGO Consortium of Taiwan, the University of Wisconsin at Milwaukee, and Lawrence Berkeley National Laboratories. Operations are conducted by COO, IPAC, and UW. 
\end{acknowledgements}

\bibliographystyle{aa}
\bibliography{37708}

\end{document}